\documentclass[10pt, a4paper]{article}
\usepackage{ictam}

\usepackage{xcolor}%made by ziqiang
\usepackage{amsmath}
\usepackage{mathabx}
\usepackage{fdsymbol}
\usepackage{tikz}

\usepackage{float}
\usepackage{graphicx}% Include figure files
\usepackage{dcolumn}% Align table columns on decimal point
\usepackage{bm}% bold math
\definecolor{mypink1}{rgb}{0.8392 0.2745 0.0353}%added by ziqiang
\definecolor{mypink2}{rgb}{0.0745 0.6235 1.0000}%added by ziqiang

\begin{document}

\title{DYNAMIC TRANSITION APPROACHING JETTING SINGULARITY DURING THE COLLAPSE OF DROP-IMPACT CRATERS}

\author[1]{\underline{Zi Qiang Yang} {\footnote{Corresponding author. E-mail: ziqiang.yang@kaust.edu.sa }}}
\author[1]{{Yuan Si Tian} } 
%Corresponding author email should be added like this
\author[1]{S. T. Thoroddsen}
\affil[1]{Division of Physical Science and Engineering,
King Abdullah University of Science and Technology (KAUST), Thuwal, Saudi Arabia}
%\affil[2]{Department of Another Author, University of Another Author, City, Country}

\maketitle

\begin{abstract}
Understanding the dynamical transition from capillary-driven to pure inertial focusing is important to revealing the mechanism underlying the most singular micro-jetting produced from drop-impact craters.  
Herein we study dimple dynamics from the collapse of these craters when the drop is of a different immiscible liquid than the pool. 
The parameter space is considerably more complex than for identical liquids, revealing intricate compound-dimple shapes. 
In contrast to the universal capillary-inertial drop pinch-off regime, where the neck radius scales as $R\sim t^{2/3}$, 
a purely inertial collapse of the air-dimple portrays $R \sim t^{1/2}$ and is sensitive to initial and boundary conditions. 
Capillary waves can therefore mold the dimple into different collapse shapes. 
We observe a clear cross-over in the nature of the dynamics from capillary-inertial to purely inertial immediately before the pinch-off.  
Notably, for singular jetting the cross-over time occurs much earlier, irrespective of the Weber number.
\end{abstract}

\section{INTRODUCTION}
%\vspace{-2.5mm}
\begin{figure}[!b]
  \centering
      \includegraphics[width=0.75\linewidth]{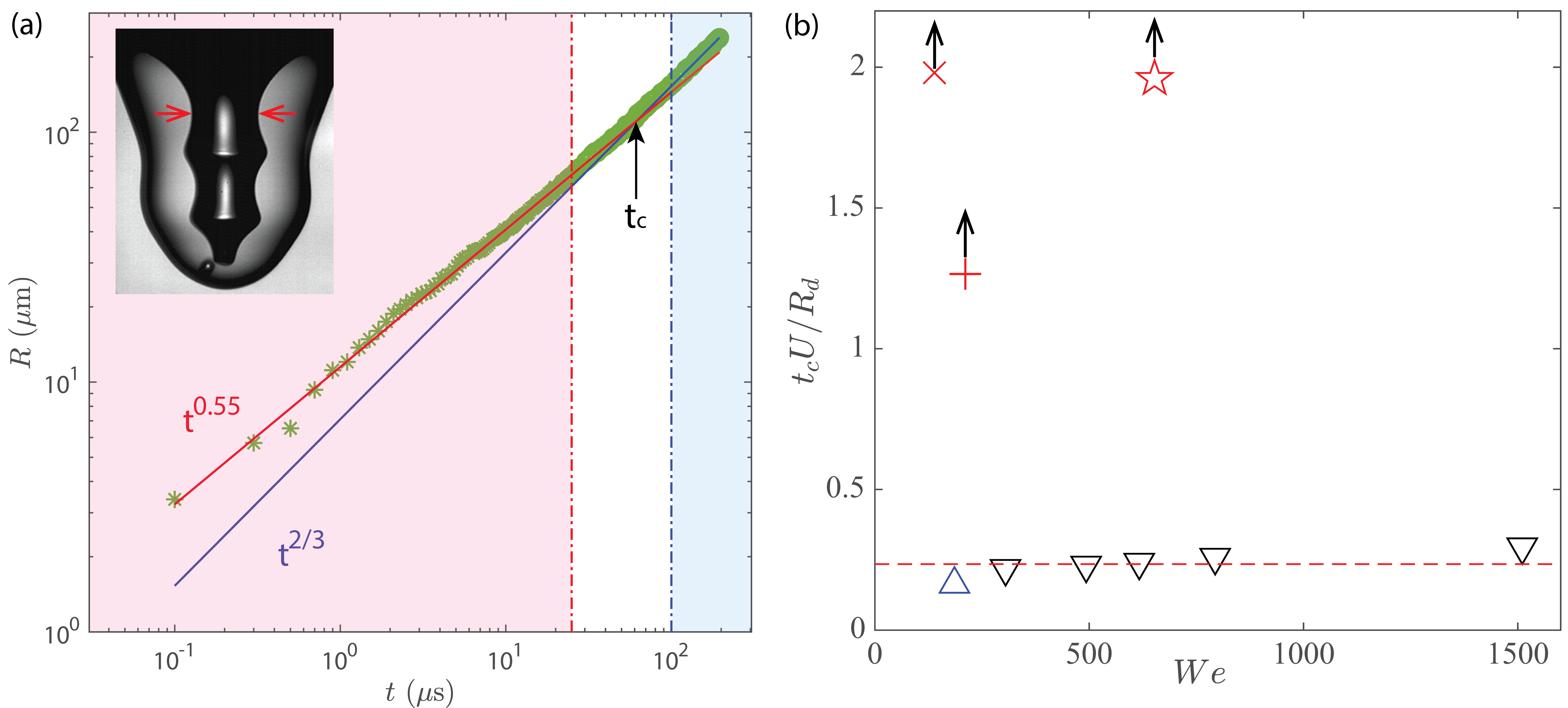}\vspace{-0.05in}\\ %figures/"Fig_2_5".eps}\vspace{-0.1in}\\
  \caption{\fontsize{9}{9}\selectfont (a) Typical dimple shapes for different impact conditions in the multi-dimple regime corresponding to the  circled red numbers in (b). Multi-pinch-offs dimple: \raisebox{.5pt}{\textcircled{\raisebox{-.9pt} {1}}}  
  $D=1.16$ mm, $U=1.7$ m/s, $Fr=259$, $We= 493$; 
  \raisebox{.5pt}{\textcircled{\raisebox{-.9pt} {2}}} $D=1.02$ mm, $U=2.1$ m/s, $Fr=421$, $We= 617$;
  \raisebox{.5pt}{\textcircled{\raisebox{-.9pt} {3}}} $D=0.93$ mm, $U=2.05$ m/s, $Fr=463$, $We= 560$ and singular telescopic dimple: \raisebox{.5pt}{\textcircled{\raisebox{-.9pt} {4}}} $D=0.73$ mm, $U=2.38$ m/s, $Fr=792$, $We= 593$.  The scale bars are 100 $\mu$m long.
  (b) Characterization of the dimples and jets in {\it Fr-We} space for drop impacts of immiscible liquids.
 The two dash curves are the bounds of the regular bubble entrapment measured by \cite{Ref8,Ref9}, for identical liquids. 
% using the best fits from \cite{oguz1990bubble}. 
 The two solid curves mark the bubble entrapment region based on our study.
% The meaning of the large symbols are given graphically in Fig. 5.
 The symbols correspond to different dimple shapes:
  (\textcolor{mypink1}{$\ovoid$}) no pinch-off shallow dimple; 
  (\textcolor{mypink2}{$\triangle$}) dimple pinch-off with bubble going out with jet; 
  (\textcolor{black}{$\triangledown$}) tiny bubble pinched off near secondary critical pinch-off;
  (\textcolor{black}{$\largewhitestar$}) singular telescopic dimple;
  (\textcolor{black}{$\triangledown$}) pinched-off bubble entrapped in PP1 drop;
  (\textcolor{blue}{$\boxvoid$}) liquid column break-up without dimple pinch-off; 
  (\textcolor{green}{$\Diamond$}) water entrapped in PP1 drop without pinch-off.
  The dashed cyan lines mark the region of multi-dimples.}
  \label{Fig_2}
\end{figure}

Singularities occur in various branches of physics and mechanics, from the gravitational collapse of black holes to the necking pinch-off of a pendent drop dripping from a faucet.  
What makes singularities a particularly attractive "laboratories" is the rapid time-scale and small length-scales
which expose the key force balance which governs the dynamics.
In fluid mechanics the pinch-off of a fluid cylinder is the most iconic example of a singularity,
which has been extensively studied over the last three decades.
There is a remarkable difference between the pinch-off of a drop from a nozzle vs that of a bubble.
For the drop the necking evolves in a self-similar manner, 
forming conical shape with capillary-inertial scaling of the necking radius, i.e. $R \sim t^{2/3}$  \cite{Ref1}.  
On the other hand, the pinch-off of a bubble follows a purely inertial dynamics, where $R \sim t^{1/2}$ \cite{Ref2,Ref3,Ref4}.
Despite a rather modest difference in the two exponent values, 
the fact that the inertia is inside the neck for the drop, while it is outside of the air-cylinder for the bubble,
changes profoundly the dynamical nature of the pinch-off.
For the purely inertia collapse the surface tension becomes irrelevant and the dynamics show strong dependence on the initial or boundary conditions.  This memory of the boundaries has been best demonstrated for a bubble pinching off from a nozzle with an elliptic cross-section \cite{Ref5,Ref6}.  

Herein, we focus on the jetting from the rebounding of a drop-impact-produced crater.  
The finest and most {\it singular jets} emerge when a dimple forms at the bottom of the crater
and collapses without pinching of a bottom bubble.
The final stage of collapse follows pure inertial scaling \cite{Ref7}.

%%%%%%%%%%%%%%%%%%%%%%%%%%%%%%%%%%%%%%%%%%%%%%

{\it EXPERIMENTAL SETUP:} 
We use different immiscible liquids: The pool is purified water, while the drop consists of PP1 (Perfluorohexane, C$_6$F$_{14}$).  PP1 is 1.71 times heavier than water and has very low surface tension $\sigma_d = 11.9$ mN/m.  The interfacial tension between the two liquids is 48 mN/m.
The range of drop sizes is $D$ = 0.6 - 2.0 mm.
By changing drop release-height we get impact velocities $U$ from 0.1 to 3.9 m/s.
The corresponding range of Reynolds, Weber and Froude numbers, based on drop properties are: 
$Re = \rho_d DU/\mu_d = 374 - 10,200$;
$We = \rho_d DU^2 /\sigma_d  = 10 - 2,000$; $Fr = U^2/(gD) = 10 - 1,500;$
where $g$ is gravity acceleration, $\rho_d$ and $\mu_d$ are drop density and dynamic viscosity.

%%%%%%%%%%%%%%%%%%%%%%%%%%%%%%%%%%%%%%%%%%%%%%%%%%%%%%%%%%%%%%%%%%%%%%%%%%%%%%%%%%%%%%%%%%%%%
%   RESULTS
%%%%%%%%%%%%%%%%%%%%%%%%%%%%%%%%%%%%%%%%%%%%%%%%%%%%%%%%%%%%%%%%%%%%%%%%%%%%%%%%%%%%%%%%%%%%%

\section{RESULTS AND DISCUSSIONS}
The impact produces a hemispheric crater into the free surface of the pool, 
with the PP1 drop liquid forming a thin continuous layer coating it.
The subsequent rebound can form a bottom dimple whose collapse produces singular jets \cite{Ref7,Ref8}.
The free surface of this dimple therefore remains between air and the PP1 drop liquid.
Figure \ref{Fig_2} highlights the regime where a dimple forms during its collapse.  
This occurs at much larger values of $We$ (based on drop properties), 
than for the classical regime (dashed lines) where the dimple entraps a bubble 
for identical liquids in both drop and pool \cite{Ref9}.
Figure \ref{Fig_2}(a) shows a prominent new feature of the dimples, 
i.e. capillary waves travelling down towards their tips, which can lead to multiple pinch-offs.  
The thinnest and fastest jets emerge when no bubble in pinched off \cite{Ref7}.

\begin{figure}[!ht]
  \centering
     \includegraphics[width=0.8\linewidth]{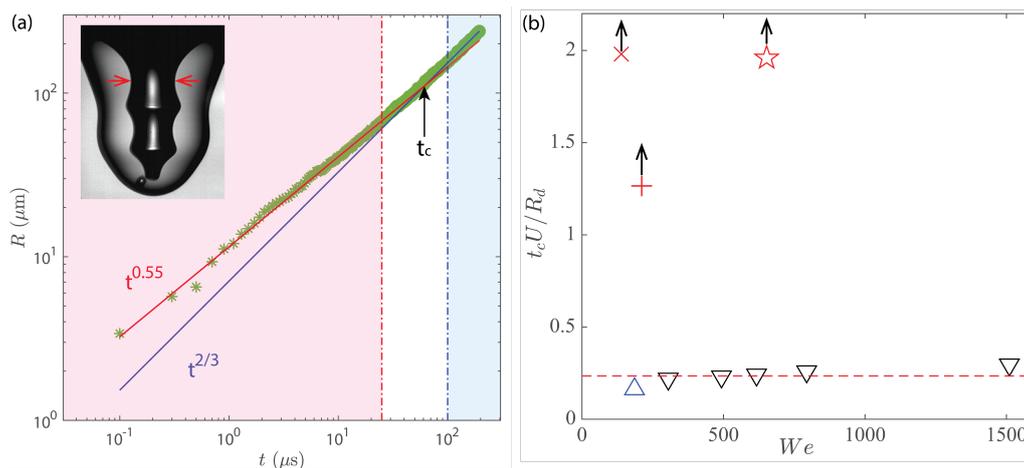}\vspace{-0.15in}\\
	\caption{\fontsize{9}{9}\selectfont (a) Logarithmic scaling of the dimple radius vs time before pinch-off. 
	There is a transition of power-law exponents from 2/3 to 0.55 closest to the pinch-off.
	The background shading marks the validity of each, with the arrow indicating the approximate cross-over time $t_c$.  
	The data is taken from two video clips spanning time-scales from 100 ns to 200 $\mu$s before pinch-off.  
	%The solid lines are the power law, black circle is experimental data and the point lines indicate where the transition starts and ends approximately. 
	%The inset shows the logarithmic scaling at the vicinity of the final pinch-off. 
	%Both the painted areas show a prominent difference of the power law.
	(b) It shows how $t_c$ normalized by the impact time $D/U$ changes with $We$, 
	for dimple pinch-off (\textcolor{mypink2}{$\triangle$} \& $\triangledown$) and singular jets (\textcolor{red}{$\times$}, \textcolor{red}{$\plus$} \& \textcolor{red}{$\largewhitestar$}).
	The vertical arrows indicate these are lower bounds, as for these cases the dynamics remain inertial for the entire video clip.}
  \label{Fig_5}
\end{figure}

What is the role of capillary waves in setting up the dimple for the inertial focusing?
For the singular jets from bursting bubbles or super-critical surface waves, 
the dimple dynamics have until recently been formulated in the self-similar capillary-inertial formalism \cite{Ref10, Ref11}, 
while the final cylindrical collapse has been shown to follow pure inertial focusing \cite{Ref7,Ref12}.
One can therefore expect a dynamical transition in the vicinity of the final jet formation.
In Figure \ref{Fig_5} we track the radius of pinch-off neck for a typical dimple, shown in the inset.
There is a clear cross-over in the nature of the dynamics from capillary-inertial $R\sim t^{2/3}$ 
to purely inertial with $R\sim t^{0.55}$ \cite{Ref3,Ref7} at $t_c \simeq 65\; \mu$s before pinch-off, as marked by the arrow.
Figure \ref{Fig_5}(b) shows that the cross-over time scales with the impact time $t_c \simeq 0.235 R_d/U$ for the pinch-off cases.
%This corresponds to only $t_c \simeq 10^{-2} /\tau_{\sigma}$, where $\tau_{\sigma}=\sqrt{\rho_d R_d^3/\sigma_d}$
%is the capillary-inertial time-scale.   % ( 1700 x 10^-9 / 0.012 )^0.5 = ( 1.7 x 10^-4 / 1.2 )^0.5 = 12 ms.
On the other hand, for singular jetting the cross-over time occurs much earlier, 
irrespective of $We$.
%with most of the pinch-off inertial $t_c/\tau_{\sigma} \sim 10^{-1}$.  
%The inset in Fig. \ref{Fig_5} shows that this is a generic property of the singular jetting, 
%%%%%%%%%%%%%%%%%%%%%%%%%%%%%%%%%%%%%%%%%%%%%%

\end{document}